\documentclass[aps,preprintnumbers,eqsecnum,amsmath,amssymb,showpacs,nofootinbib]{revtex4}
\usepackage{graphicx}
\usepackage{dcolumn}\usepackage{bm}\usepackage{epsfig}

\begin{document} % HEPHY-PUB 925/13
\title{\boldmath Accurate bottom-quark mass from Borel QCD sum rules for $f_B$ and $f_{B_s}$}
\author{Wolfgang Lucha$^{a}$, Dmitri Melikhov$^{a,b,c}$, and Silvano Simula$^{d}$}
\affiliation{$^a$HEPHY, Austrian Academy of Sciences, Nikolsdorfergasse 18, A-1050, Vienna, Austria
\\
$^b$SINP, Moscow State University, 119991, Moscow, Russia
\\
$^c$Faculty of Physics, University of Vienna, Boltzmanngasse 5, A-1090 Vienna, Austria
\\
$^d$INFN, Sezione di Roma Tre, Via della Vasca Navale 84, I-00146,
Roma, Italy}
\begin{abstract}
We prove that Borel QCD sum rules for
heavy--light currents yield very strong correlations between the
$b$-quark mass $m_b$ and the $B$-meson decay constant $f_B,$
namely, $\delta f_B/f_B\approx-8\,\delta m_b/m_b.$ This fact opens
the possibility of an accurate sum-rule extraction of $m_b$ by
using $f_B$ as input. Combining precise lattice QCD
determinations of $f_B$ with our sum-rule analysis based on the
three-loop $O(\alpha_s^2)$ heavy--light correlation function leads
to $\overline{m}_b(\overline{m}_b)=(4.247\pm0.034)\;\mbox{GeV}.$
\end{abstract}
\maketitle

\section{Introduction}
The $b$-quark mass --- for instance, the
$\overline{\rm MS}$ running mass at renormalization scale $\nu,$
$\overline{m}_b(\nu),$ or
$m_b\equiv\overline{m}_b(\overline{m}_b)$ ---~is~one of the
fundamental parameters of the standard model and therefore its
precise knowledge is highly desirable. The latest edition of the Review of Particle Physics reports 
$m_b=(4.18\pm 0.03)\;\mbox{GeV}$ \cite{pdg}. 

A direct way to determine $m_b$ is by means of lattice QCD simulations; however, since
the physical $b$-quark is too heavy for current lattice setups,
the determination of $m_b$ from pure lattice QCD requires either
the extrapolation of the lattice results from lighter simulated
masses or the use of the heavy-quark effective theory (HQET)
formulated on the lattice. Using the former approach, the values
$m_b=(4.29\pm0.14)\;\mbox{GeV}$ \cite{ETMC1} and
$m_b=(4.35\pm0.12)\;\mbox{GeV}$ \cite{ETMC2} have recently been
deduced, while the results $m_b=(4.26\pm0.09)\;\mbox{GeV}$
\cite{mb_Gimenez}, $m_b=(4.25\pm0.11)\;\mbox{GeV}$
\cite{mb_UKQCD}, and $m_b=(4.22\pm0.11)\;\mbox{GeV}$
\cite{ALPHA}~have been determined adopting the HQET-based
approach. All above findings have been obtained using unquenched
gauge configurations with $N_f=2$ dynamical flavors in the sea.

Recently, more accurate determinations of the $b$-quark mass have
been performed by exploiting moment sum rules for two-point
functions of heavy--heavy currents: low-$n$ moment sum rules based
on three-loop $O(\alpha_s^2)$ \cite{mb4} and four-loop
$O(\alpha_s^3)$ \cite{mb1} fixed-order pQCD calculations combined
with the experimental data yield $m_b=(4.209\pm0.050)\;\mbox{GeV}$
\cite{mb4} and $m_b=(4.163\pm0.016)\;\mbox{GeV}$ \cite{mb1},
respectively. The latter finding has been confirmed by a study
based on a combination of perturbative QCD and lattice QCD
simulations with $N_f=2+1$ dynamical flavors in the sea
\cite{mb2}. Combining~large-$n$ moments obtained within
renormalization-group-improved NNLL-order $\Upsilon$ sum rules
with the experimental data yields
$m_b=(4.235\pm0.055_{\rm(pert)}\pm0.03_{\rm(exp)})\;\mbox{GeV}$
\cite{hoang}.

In this paper, we show that the Borel QCD sum rules for
heavy--light correlators provide the possibility to extract~the
bottom-quark mass with comparable accuracy if a precise value for
the $B$-meson decay constant $f_B$ is adopted as~input.

Let us first explore what degree of sensitivity of $f_B$ to the
precise value of the $b$-quark mass can be expected on the basis
of simple quantum-mechanical considerations. Nonrelativistic
potential models predict the following relationship between the
ground-state wave function at the origin, $\psi(r=0),$ and the
ground-state binding energy $\varepsilon$:
\begin{eqnarray}
\label{qm1}
|\psi(r=0)|\propto\varepsilon^{3/2}.
\end{eqnarray}
Equation (\ref{qm1}) is exact for any ground state in a purely
Coulomb or purely harmonic-oscillator potential (or, to be more
precise, in any model where the potential involves only one
coupling constant). Moreover, this relation proves to be a good
approximation for ground states in potentials given by the sum of
confining and Coulomb interactions~\cite{lms_qcdvsqm}.

Now, taking into account that the decay constant is the analogue
of the wave function at the origin and incorporating the known
scaling behaviour of the decay constant of a heavy meson in the
heavy-quark limit \cite{neubert} --- which should~work well for
the beauty mesons --- one obtains an approximate relation between
the $B$-meson mass $M_B$ and the heavy-quark pole mass $m_Q$:
\begin{eqnarray}
\label{qm2}
f_B\sqrt{M_B}=\kappa(M_B-m_Q)^{3/2}.
\end{eqnarray}

Keeping the ground-state mass fixed and equal to its experimental
value $M_B=5.27\;\mbox{GeV}$, we can easily pin down the
dependence of $f_B$ on small variations $\delta m_Q$ of the
heavy-quark pole mass near some average value of $m_Q$. Taking
into account that $f_B\approx200\;\mbox{MeV}$ for
$m_Q\approx4.6\div4.7\;\mbox{GeV},$ we end up with
$\kappa\approx0.9\div1.0$ and $\delta f_B\approx-0.5\,\delta
m_Q$~or,~equivalently,
\begin{eqnarray}
\label{qm3}\frac{\delta f_B}{f_B}\approx-(11\div12)\,\frac{\delta m_Q}{m_Q}.
\end{eqnarray}
Thus, the sensitivity of $f_B$ to the precise value of the
heavy-quark mass should be rather high: a variation of~the~quark
mass by $+100\;\mbox{MeV}$ entails $\delta
f_B\approx-50\;\mbox{MeV}.$ Clearly, a similar effect should be
observable in the outcomes of QCD~sum rules \cite{svz,aliev}.

Recently, several QCD sum-rule analyses
\cite{nar2001,jamin,lms_fB,nar2012} of beauty-meson decay
constants relying on three-loop heavy--light correlators
\cite{chetyrkin} have been published, see Table \ref{Table:1}
(note that all results collected in Table \ref{Table:1} are
obtained by applying~the QCD sum-rule method to essentially the
same analytical expression for the correlator).

\begin{table}[!ht]
\caption{Some recent QCD sum-rule predictions for $f_B$ from heavy--light two-point functions.}
\label{Table:1}
\vspace{2ex}
\begin{tabular}{ccccc}\hline\hline
&Ref.~\cite{nar2001}&Ref.~\cite{jamin}&Ref.~\cite{lms_fB}&Ref.~\cite{nar2012}\\\hline
$m_b$~(GeV)&$4.05\pm0.06$&$4.21\pm0.05$&$4.245\pm0.025$& $4.236\pm0.069$\\
$f_B$~(MeV)&$203\pm23$&$210\pm19$&$193\pm15$ & $206\pm 7$\\\hline\hline\end{tabular}
\end{table}

At first glance, the QCD sum-rule results for $f_B$ seem to be
very stable and practically independent of the input~value of
$m_b$. This, however, may not be regarded as an argument in favour
of the published predictions: Obviously, the~results in Table
\ref{Table:1} do not follow the general pattern discussed above;
for instance, the central values of $m_b$ reported in
\cite{nar2001}~and \cite{nar2012} differ by some $200\;\mbox{MeV},$ 
whereas the corresponding decay constants remain almost unchanged. 
Therefore, we are forced to conclude that
not all the results in Table \ref{Table:1} are equally trustable.

Recall that the values of the ground-state parameters in
Table \ref{Table:1} are strongly influenced by (i) the way~one
reorganizes the three-loop perturbative result in terms of the
pole or the running mass of the heavy quark, and (ii) by one's way
of fixing the auxiliary parameters of the sum-rule approach,
particularly, the effective continuum threshold.

The goal of this paper is to present a critical detailed analysis
of the sum-rule extraction of $f_B.$ Our main conclusion is that
if the appropriate expression for the correlator in terms of the
running heavy-quark mass is used and consistent procedures for
extracting the bound-state parameters are applied, the QCD
sum-rule results are in excellent agreement with the behavior
expected from quantum mechanics: the decay constant $f_B$ obtained
from QCD sum rules is strongly correlated with the input value of
the heavy-quark mass $m_b.$ For all other input parameters of the
correlator (quark condensate, $\alpha_s,$ renormalization scale,
etc.) fixed, we find
\begin{eqnarray}
f_B(m_b)=\left(192.0-37\,\frac{m_b-4.247\;\mbox{GeV}}{0.1\;\mbox{GeV}}
\pm3_{\rm(syst)}\right)\mbox{MeV}.
\end{eqnarray}
Evidently, the dependence of $f_B$ on $m_b$ agrees very well with
the semi-qualitative quantum-mechanical expression~(\ref{qm3}).
The strong correlation between $f_B$ and $m_b$ opens the
possibility to deduce an accurate value of $m_b$ using $f_B$
as~input. Combining our sum-rule analysis based on the
heavy--light correlator known to order $\alpha_s^2$ with the
average of the most recent determinations of $f_B$ from lattice
QCD, $f_B^{\rm LQCD}=(191.5\pm7.3)\;\mbox{MeV}$ (see Table
\ref{table:lattice}), leads to the accurate~estimate
\begin{eqnarray}
m_b=(4.247\pm0.034)\;\mbox{GeV}.
\end{eqnarray}

\vspace{-.2cm}
\begin{table}[h]
\caption{Some recent lattice-QCD evaluations of $f_B$ and $f_{B_s}$.}
\label{table:lattice}
\vspace{2ex}
\begin{tabular}{ccccc}\hline\hline Collaboration&$N_f$&$f_B$
(MeV)&$f_{B_s}$ (MeV)&$f_{B_s}/f_B$
\\
\hline
ETMC I
\cite{ETMC1}
&2&$195\pm12$&$232\pm10$&$1.19\pm0.05$
\\
ETMC II
\cite{ETMC2}&2&$197\pm10$&$234\pm6$&$1.19\pm0.05$
\\
ALPHA
\cite{ALPHA}&2&$193\pm10$&$219\pm12$&$1.13\pm0.09$
\\
\hline
HPQCD I
\cite{HPQCD1}&2+1&$191\pm9$&$228\pm10$&$1.188\pm0.018$
\\HPQCD
II \cite{HPQCD2}&2+1&$189\pm4$&$225\pm4$&---
\\FNAL/MILC
\cite{MILC}&2+1&$196.9\pm9.1$&$242\pm10$&$1.229\pm0.026$
\\
\hline
our average&&$191.5\pm7.3$&$228.8\pm6.9$&$1.198\pm0.030$
\\\hline\hline
\end{tabular}
\end{table}
This paper is organized as follows: In the next section, we
discuss the convergence of the OPE series for the~correlator
expressed in terms of either pole or running quark mass. Section
\ref{Sec:extr} presents the details of the extraction
procedure~with particular emphasis on the related uncertainties of
the extracted parameters. Section \ref{Sec:conc} summarizes our
conclusions.

%************************************************************************************
\section{Correlation function, operator product expansion, and heavy-quark mass}
The basic object for our study of the decay constants of heavy
pseudoscalar $B$ (or $B_s$) mesons is the
correlator~\cite{svz,aliev}
\begin{eqnarray}\label{1.1}\Pi(p^2)=i\int
d^4x\,e^{ipx}\left\langle0\left|T\left(j_5(x)j^\dag_5(0)\right)\right|0\right\rangle
\end{eqnarray}
of two pseudoscalar heavy--light currents
\begin{eqnarray}
j_5(x)=(m_b+m)\,\bar q(x)i\gamma_5b(x),
\end{eqnarray}
where $q(x)$ denotes the field of the light quark of mass $m,$
that is, $q(x)\equiv d(x)$ for $B$ and $q(x)\equiv s(x)$ for
$B_s$. The OPE for this correlator may be calculated by using
perturbative QCD and adding nonperturbative power corrections
given in terms of vacuum condensates. The QCD sum rule for this
correlator is obtained by equating the Borelized OPE~for the
correlator (\ref{1.1}), $\Pi(p^2)\to\Pi(\tau),$ and the Borelized
correlator calculated by insertion of intermediate hadron states:
\begin{eqnarray}
\label{pitau}
\Pi(\tau)=f_B^2 M_B^4 e^{-M_B^2\tau}+\int\limits_{s_{\rm phys}}^\infty ds\,e^{-s\tau}\rho_{\rm hadr}(s)=
\int\limits_{(m_b+m)^2}^\infty ds\,e^{-s\tau}\rho_{\rm pert}(s,\mu)+\Pi_{\rm power}(\tau,\mu),
\end{eqnarray}
where $M_B$ is the mass of the $B$ (or $B_s$) meson and $f_B$ is
the decay constant of the $B$ (or $B_s$) meson, defined by
\begin{eqnarray}\label{decay_constant}
(m_b+m)\langle0|\bar qi\gamma_5b|B\rangle=f_BM_B^2.
\end{eqnarray}
In (\ref{pitau}), $s_{\rm phys}=(M_{B^*}+M_P)^2$ is the physical
continuum threshold, fixed by the mass $M_{B^*}$ of the beauty
vector~meson and the mass $M_P$ of the lightest pseudoscalar with
appropriate quantum numbers (the pion or the kaon, respectively).

For large values of $\tau,$ the contributions of excited states to
the Borelized correlator (\ref{pitau}) decrease faster than the
ground-state contribution and thus $\Pi(\tau)$ is saturated by the
ground state. Therefore, knowing the correlator at large~$\tau$
provides direct access to the ground-state parameters. However,
analytic results for the correlator are obtained from~a truncated
OPE, which yields a good approximation to the correlator only at
not too large $\tau,$ where excited states~still contribute
sizably to $\Pi(\tau).$

To exclude the excited-state contributions from the sum rule
(\ref{pitau}), one adopts the duality Ansatz: all contributions of
excited states are counterbalanced by the perturbative
contribution above an {\em effective continuum threshold},~$s_{\rm eff}(\tau),$
which differs from the physical continuum threshold.
While the physical continuum threshold is a constant determined by
the masses of the lightest hadrons that may be produced from the
vacuum by the interpolating current, the effective continuum
threshold is a parameter of the sum-rule approach. The effective
continuum threshold has interesting and nontrivial properties
which have been discussed in great detail in \cite{lms_1}. In
particular, it has been demonstrated that the ``true'' effective
threshold which correctly reproduces the true ground-state
parameters is a $\tau$-dependent function~\cite{lms_new}.

Applying the duality assumption entails the following relation
between the ground-state contribution and the OPE:
\begin{eqnarray}
\label{sr}
f_B^2 M_B^4 e^{-M_B^2\tau}=\int\limits_{(m_b+m)^2}^{s_{\rm eff}(\tau)}ds\,e^{-s\tau}\rho_{\rm pert}(s,\mu)+
\Pi_{\rm power}(\tau,\mu)\equiv\Pi_{\rm dual}(\tau,s_{\rm eff}(\tau)).
\end{eqnarray}
We refer to the right-hand side of this equation as the {\em dual
correlator} $\Pi_{\rm dual}(\tau,s_{\rm eff}(\tau)).$

Clearly, even if the QCD inputs $\rho_{\rm pert}(s,\mu)$ and
$\Pi_{\rm power}(\tau,\mu)$ are known, the extraction of the decay
constant requires, in addition, a criterion for determining
$s_{\rm eff}(\tau)$. As first step, however, we need a reasonably
convergent OPE for both correlator and dual correlator. For
heavy--light systems, the relative sizes of the lowest-order
terms, which contain powers of the heavy-quark mass, turn out to
depend strongly on one's choice of the renormalization scheme and
scale.

The best-known three-loop calculations of the perturbative
spectral density \cite{chetyrkin} have been performed in form of
an expansion in terms of the $\overline{\rm MS}$ strong coupling
$\alpha_{\rm s}(\mu)$ and the pole mass of the heavy-quark $M_b$:
\begin{eqnarray}
\label{rhopert}
\rho_{\rm pert}(s,\mu)
=\rho^{(0)}(s,M_b^2)+\frac{\alpha_{\rm s}(\mu)}{\pi}\rho^{(1)}(s,M_b^2)+
\left(\frac{\alpha_{\rm s}(\mu)}{\pi}\right)^2\rho^{(2)}(s,M_b^2,\mu)+\cdots.
\end{eqnarray}
The correlator (\ref{1.1}) and its Borel image (\ref{pitau}) do
not depend on the renormalization scale $\mu$. Unfortunately, this
nice property is lost if one works with truncated expansions: both
the perturbative expansion truncated at fixed order in
$\alpha_{\rm s}$ and the lowest-order power corrections $\Pi_{\rm
power}(\tau,\mu)$ given in terms of condensates and the radiative
corrections~to~the latter depend on $\mu$.

\begin{figure}[ht]
\begin{tabular}{cc}
\includegraphics[width=8.5cm]{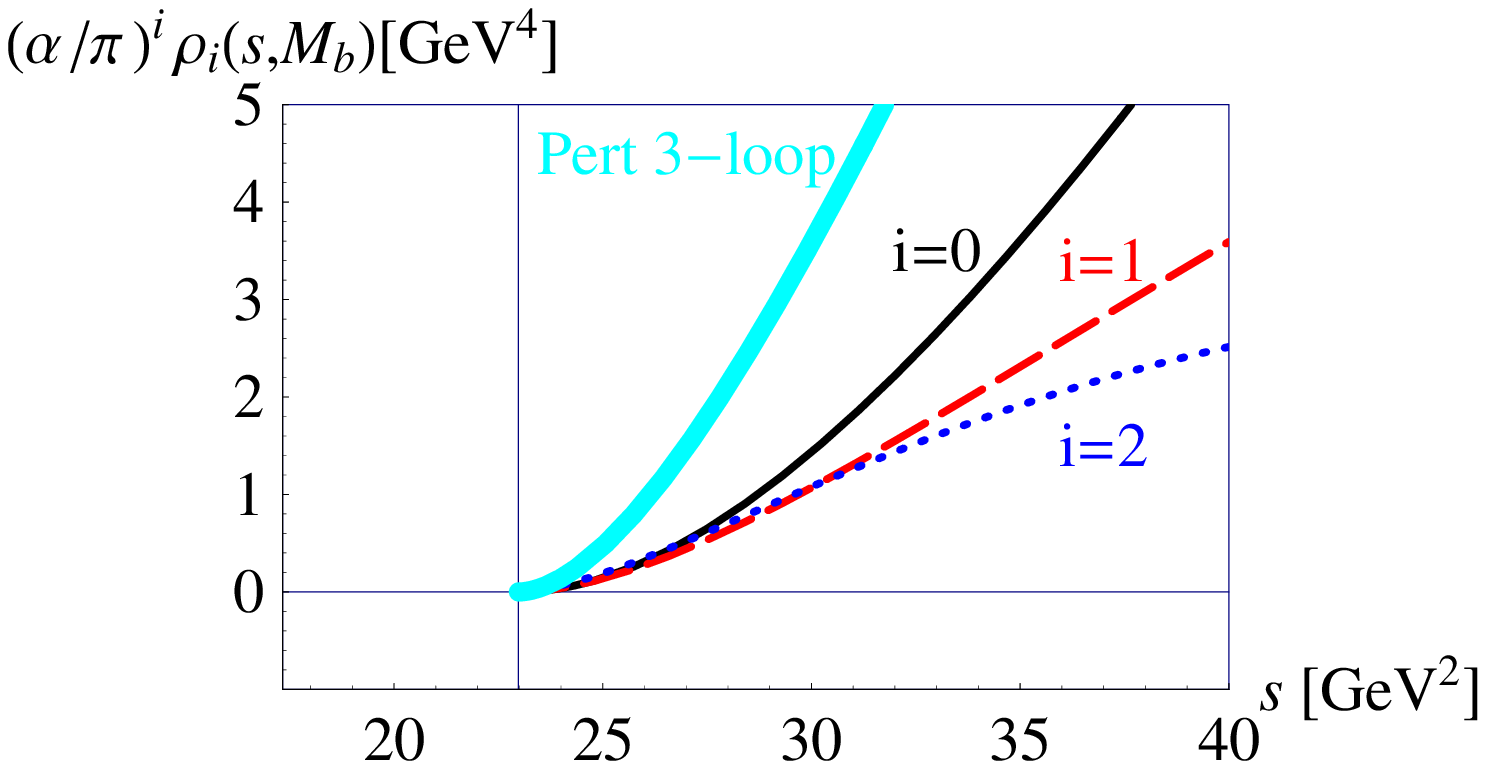}&
\includegraphics[width=8.5cm]{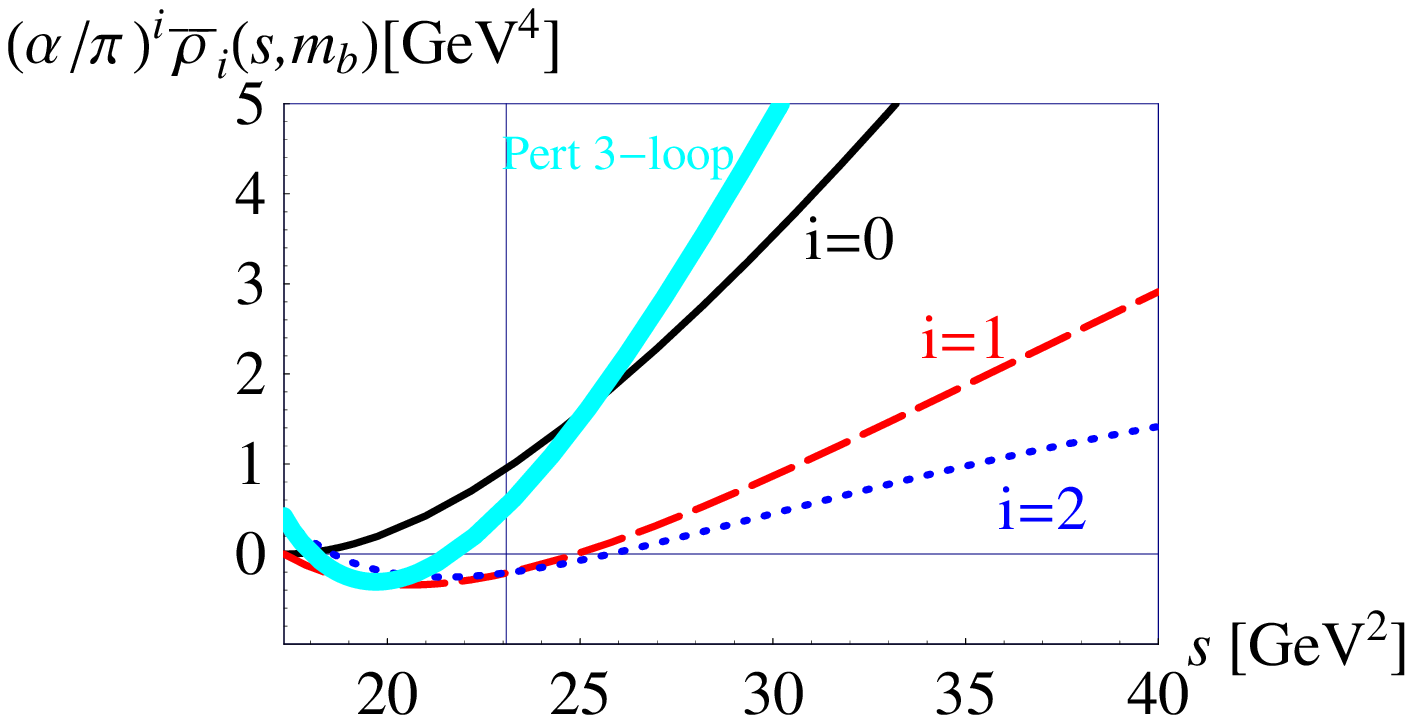}\\
\includegraphics[width=8.5cm]{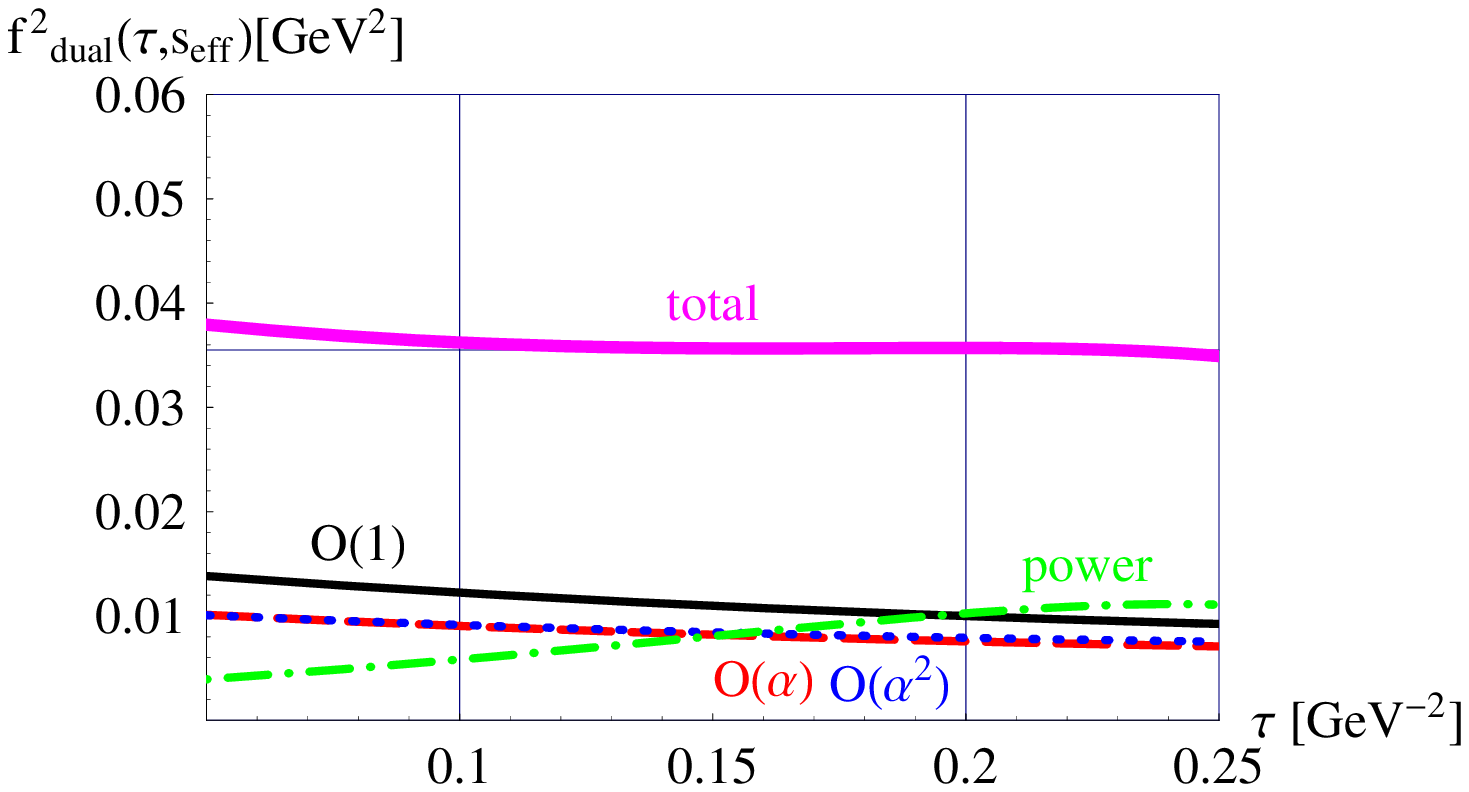}&
\includegraphics[width=8.5cm]{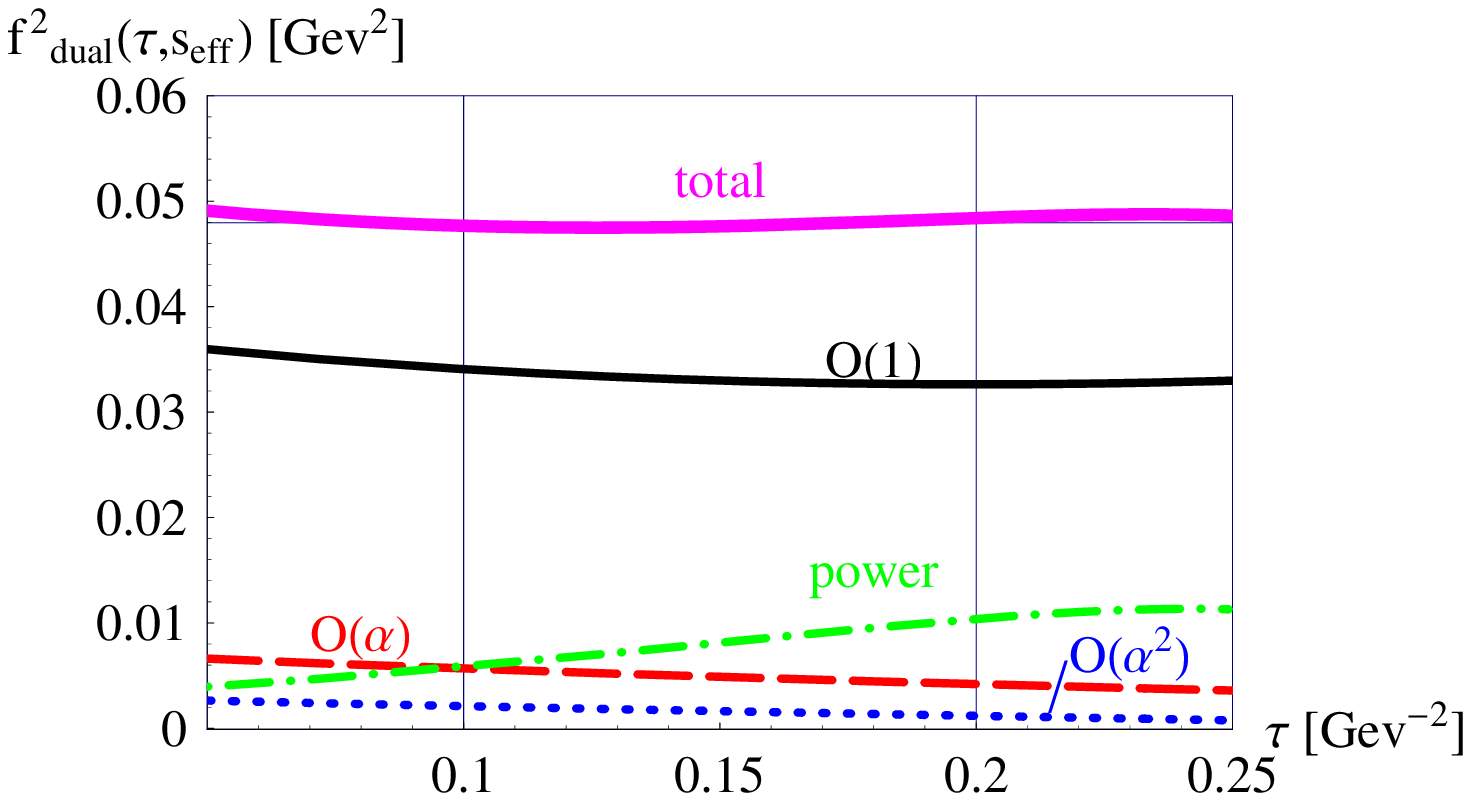}
\end{tabular}
\caption{OPE computed in terms of pole mass (left) and
$\overline{\rm MS}$ mass (right) of the $b$-quark. First row:
spectral densities; second~row: corresponding sum-rule findings
for $f_B$. In both cases, a typical value of the effective
continuum threshold is used: $s_0=35\;\mbox{GeV}^2.$ Bold solid
lines: total result; black solid lines: $O(1)$ contribution; red
dashed lines: $O(\alpha_s)$ contribution; blue dotted
lines:~$O(\alpha_s^2)$ contribution; green dash-dotted lines:
power contributions.}
\label{Plot:1}
\end{figure}

The pole mass has been used in most sum-rule studies since the
pioneering work \cite{aliev}. It turns out, however, that~the OPE
for the dual correlator expressed in terms of the pole mass $M_b$
exhibits a bad convergence, as illustrated in Fig.~\ref{Plot:1}.

An alternative option \cite{jamin} is to reorganize the
perturbative expansion in terms of the running $\overline{\rm MS}$
mass, $\overline{m}_b(\nu)$,~by substituting $M_b$ in the spectral
densities $\rho^{(i)}(s,M_b^2)$ via its perturbative expansion in
terms of the running mass $\overline{m}_b(\nu)$ (while keeping the
integration variable $s$ fixed)\footnote{Note that the scale $\mu$
of $\alpha_s(\mu)$ in the expansion (\ref{rhopert}) is not
necessarily equal to the scale $\nu$ in the relation (\ref{2.7})
between $M_b$ and~$\overline{m}_b(\nu)$;~the two scales are, in
principle, independent. As noticed in \cite{hoang}, setting these
scales equal to each other leads to a reduced dependence of the
truncated correlator on the then common scale; however, more
realistic error estimates are obtained if one studies the
sensitivity of the truncated correlator to independent variations
of the scales $\mu$ and $\nu$.}
\begin{eqnarray}
\label{2.7}
M_b=\overline{m}_b(\nu)\left(1+\frac{\alpha_s(\nu)}{\pi}\,r_1
+\left(\frac{\alpha_s(\nu)}{\pi}\right)^2r_2+\ldots \right).
\end{eqnarray}
Since the original correlator is known to order $\alpha_s^2,$ it
suffices to use the relation between $M_b$ and
$\overline{m}_b(\mu)$ to $\alpha_s^2$ accuracy (although the
$\alpha_s^3$ term is also known \cite{Mb3loop}). Moreover, one
should omit all terms of order $\alpha_s^3$ and higher,
induced~by~the substitution $M_b\to \overline{m}_b(\nu)$ in the
results of \cite{chetyrkin}. Explicit expressions for the
perturbative spectral densities and power corrections may be found
in \cite{chetyrkin,jamin} and are not given here. Notice that two
different scales, $\mu$ and $\nu$, naturally emerge when
reorganizing the perturbative expansion from the pole $b$-quark
mass to the running $b$-quark mass. One may set these scales equal
to each other; we, however, leave the scales independent from each
other and investigate the impact of particular choices of the
scales $\mu$ and $\nu$ on the extracted values of $m_b$ and the
decay constants $f_B$ and~$f_{Bs}$.

One caveat is in order here: the spectral density (\ref{rhopert})
involves an implicit $\theta$-function restricting the
integration~region in the correlator: for instance, for a massless
light quark, it reads $\theta(s-M_b^2)$. Switching from the pole
to the running mass, $M_b\to\overline{m}_b(\nu)$, this
$\theta$-function has to be expanded in powers of $\alpha_s,$
which step induces ``surface'' terms $\delta(s-M_b^2)$ and their
derivatives. The spectral densities $\rho^{(i)}(s,M_b^2),$
however, have zeros of second order at this threshold $s=M_b^2;$
consequently, to the $O(\alpha_s^2)$ accuracy considered, the
surface terms do not contribute and one merely has to perform~the
replacement $\theta(s-M_b^2)\to \theta(s-\overline{m}_b^2(\nu))$.
The surface terms enter the game at order $\alpha_s^3$ and higher.

In order to appreciate the amount of improvement achieved by
reorganizing the perturbative expansion in terms of the running
mass, Fig.~\ref{Plot:1} shows the perturbative spectral densities
and the estimates for $f_B$ arising from the sum~rule (\ref{sr})
for two choices of the $b$-quark mass: the pole mass $M_b$ and the
running $\overline{\rm MS}$ mass $\overline{m}_b(\nu)$. All
results are given~for $m_b=4.163\;\mbox{GeV},$ corresponding to
2-loop and 3-loop pole masses $M_b^{\rm 2-loop}=4.75\;\mbox{GeV}$
and $M_b^{\rm 3-loop}=4.89\;\mbox{GeV}$~\cite{Mb3loop}. Since we
work at $O(\alpha_s^2)$ accuracy, we use for consistency the
2-loop value of $M_b$ to obtain the results depicted
in~Fig.~\ref{Plot:1}. For the other relevant OPE parameters, we
adopt the following values \cite{pdg,colangelo}:
\begin{align}
\label{Table:2}
&m_d(2\;\mbox{GeV})=(3.5\pm0.5)\;\mbox{MeV},\quad
m_s(2\;\mbox{GeV})=(95\pm5)\;\mbox{MeV},\quad
\alpha_s(M_Z)=0.1184\pm0.0007,\\ &\langle\bar
qq\rangle(2\;\mbox{GeV})=-((269\pm17)\;\mbox{MeV})^3,\quad
\langle\bar ss\rangle(2\;\mbox{GeV})/\langle\bar
qq\rangle(2\;\mbox{GeV})=0.8\pm0.3,
\quad\left\langle\frac{\alpha_s}{\pi}\,GG\right\rangle
=(0.024\pm0.012)\;\mbox{GeV}^4.
\nonumber
\end{align}
The sum-rule estimates shown in Fig.~\ref{Plot:1} are obtained for
$\mu=\nu=m_b$ and for a $\tau$-independent effective
threshold~$s_{\rm eff}$. Clearly, the choice of the heavy-quark
mass (that is, pole or running) used in the OPE makes a great
difference for~the numerical values of the truncated heavy-light
correlators and of the resulting decay constants.

The above observations may be summarized as follows:
\begin{enumerate}
\item When the dual correlator is calculated in
terms of the heavy-quark pole mass, its perturbative expansion
exhibits no sign of convergence; the contributions of the $O(1),$
$O(\alpha_s),$ and $O(\alpha_s^2)$ terms are of nearly the same
magnitude. Therefore, in this scheme one cannot expect higher
orders to give smaller contributions.\item Formulating the
perturbative series in terms of the heavy-quark $\overline{\rm
MS}$ mass yields a clear hierarchy of contributions.\item The
decay constant extracted from the pole-mass truncated OPE
($f_B=188\;\mbox{MeV}$) is substantially smaller~than that from
the $\overline{\rm MS}$-mass OPE truncated at the same order
($f_B=220\;\mbox{MeV}$). Nevertheless, both decay constants
exhibit a satisfactory degree of stability over a wide range of
the Borel parameter! We therefore stress again~that mere Borel
stability is by far not sufficient to guarantee the reliability of
any sum-rule extraction of bound-state feature. We have
illustrated these findings before in some exactly solvable
quantum-mechanical examples~\cite{lms_1,lms_qcdvsqm}.
\end{enumerate}
Because of the evident lack of convergence of the truncated
pole-mass OPE for the correlator, we employ the $\overline{\rm
MS}$-mass OPE in our subsequent sum-rule analysis.

\section{Extraction of the decay constant}
\label{Sec:extr}
According to the standard procedures of the QCD sum-rule approach,
its application requires the following steps:

\subsubsection{The Borel window}The working $\tau$-window is chosen
such that the OPE gives a sufficiently accurate description~of the
exact correlator (i.e., all higher-order radiative and power
corrections are small) and at the same time the ground state gives
a ``sizable'' contribution to the correlator. Hence, we require
\cite{lms_new,lms_qcdvsqm,lms_fD} that the power corrections do
not exceed 30\% of the dual correlator (to fix the maximal $\tau$)
and that the ground-state contribution does not fall below 10\%
(to fix the~minimal~$\tau$).

In practice, our $\tau$-window for the $B_{(s)}$ mesons is
$0.05\lesssim\tau\,\mbox{(GeV$^{-2}$)}\lesssim0.175$. Such a
window is much more extended than the $\tau$-range usually adopted
in the literature, e.g.,
$0.17\lesssim\tau\,\mbox{(GeV$^{-2}$)}\lesssim0.25$ \cite{jamin}
or
$0.20\lesssim\tau\,\mbox{(GeV$^{-2}$)}\lesssim0.26$~\cite{nar2012}.
We observe (i) that our upper bound in $\tau$ is much safer with
respect to the convergence properties of both perturbative and
power correction series, and (ii) that our lower bound in $\tau$
produces a dual correlator (\ref{sr}) which represents the
ground-state contribution in a much wider range of values of
$\tau$. The latter property corresponds to the fact that the
quantity $\left[\Pi_{\rm dual}(\tau,s_{\rm eff}(\tau))\cdot
e^{M_B^2\tau}\right]$ should exhibit a plateau in a wide range of
values of $\tau$, which makes the extraction~of the decay constant
$f_B$ much more reliable.

Finally, we notice that it would be extremely unreasonable to
assume a $\tau$-independent effective threshold $s_{\rm eff}$ in a
$\tau$-window where the impact of the contamination of excited
states in the full correlator changes quite significantly, as
explicitly shown in Ref.~\cite{lms_new}.

\subsubsection{The effective continuum threshold}
To find $s_{\rm eff}(\tau),$ we employ a previously developed
algorithm \cite{lms_new,lms_qcdvsqm}, which has proven to provide
a reliable extraction of the ground-state parameters in
quantum-mechanical models and of the charmed-meson decay constants
in QCD~\cite{lms_fD}. We introduce the {\em dual invariant mass\/}
$M_{\rm dual}$ and the {\em dual decay constant\/} $f_{\rm dual}$
by the definitions
\begin{eqnarray}
\label{dual}
M_{\rm dual}^2(\tau)\equiv-\frac{d}{d\tau}\log\Pi_{\rm
dual}(\tau,s_{\rm eff}(\tau)),\qquad f_{\rm dual}^2(\tau)\equiv
M_B^{-4}\,e^{M_B^2\tau}\,\Pi_{\rm dual}(\tau,s_{\rm eff}(\tau)).
\end{eqnarray}
By construction, the dual mass should reproduce the true
ground-state mass $M_B.$ So, the deviation of $M_{\rm dual}$
from~$M_B$ measures the contamination of the dual correlator by
excited states. Starting from an Ansatz for $s_{\rm eff}(\tau)$
and requiring a minimum deviation of $M_{\rm dual}$ from $M_B$ in
the $\tau$-window generates a variational solution for $s_{\rm
eff}(\tau).$ With the latter at our disposal, $f_{\rm dual}(\tau)$
yields the desired decay-constant estimate. Since we deal with a
limited $\tau$-window, it suffices to consider polynomials in
$\tau$, including also the standard assumption for the effective
threshold, a $\tau$-independent~constant:
\begin{eqnarray}
\label{zeff}
s^{(n)}_{\rm eff}(\tau)=\sum\limits_{j=0}^ns_j^{(n)}\tau^{j}.
\end{eqnarray}
We obtain the expansion coefficients $s_j^{(n)}$ by minimizing the
squared difference between $M^2_{\rm dual}$ and $M^2_B$ in the
$\tau$-window:
\begin{eqnarray}
\label{chisq}
\chi^2\equiv\frac{1}{N}\sum_{i=1}^N\left[M^2_{\rm dual}(\tau_i)-M_B^2\right]^2.
\end{eqnarray}

\vspace{-.5cm}
\subsubsection{Uncertainties in the extracted decay constant}
The resulting value of the decay constant $f_{B_{(s)}}$ is, beyond
doubt, sensitive to the input values of the OPE parameters ---
which determines what we call the {\em OPE-related error\/} ---
and to the details of the adopted prescription for fixing~the
behaviour of the effective continuum threshold $s_{\rm eff}(\tau)$
--- which we will refer to as the {\em systematic error}.

\vspace{1.5ex}\noindent{\em OPE-related error:} We estimate the
size of the OPE-related error by perform a bootstrap analysis
\cite{bootstrap}, allowing the OPE parameters to vary over the
ranges indicated in (\ref{Table:2}) and using 1000 bootstrap
events. Gaussian distributions~for all OPE parameters but the
scales $\mu$ and $\nu$ are employed. For the latter, we assume
uniform distributions in the range $3 \leq \mu,\nu\;(\mathrm{GeV})
\leq 6$. The resulting distribution of the decay constant turns
out to be close to Gaussian shape. Hence, the quoted OPE-related
error is a Gaussian error.

\vspace{1.5ex}\noindent{\em Systematic error:} The systematic
error, encoding the limited intrinsic accuracy of the sum-rule
method, constitutes a rather subtle point. In quantum mechanics,
we observed, for polynomial parameterizations of the effective
continuum threshold $s_{\rm eff}(\tau),$ that the band of results
obtained from linear, quadratic,~and cubic Ans\"atze for the
effective threshold encompasses the true value of the decay
constant \cite{lms_new}. Moreover, the extraction procedures in
quantum mechanics~and in QCD proved to be strikingly similar
\cite{lms_qcdvsqm}. Thus, the half-width of this band may be
regarded as a realistic estimate~for the systematic uncertainty of
the prediction. The ultimate efficiency and reliability of this
algorithm has already been established for the decay constants of
$D$ and $D_s$ mesons \cite{lms_fD}. Here, we apply this technique
to the $B$ and $B_s$ mesons.

\subsection{Decay constant of the $B$ meson}
\begin{figure}[ht]
\begin{tabular}{cc}
\includegraphics[width=6.7cm]{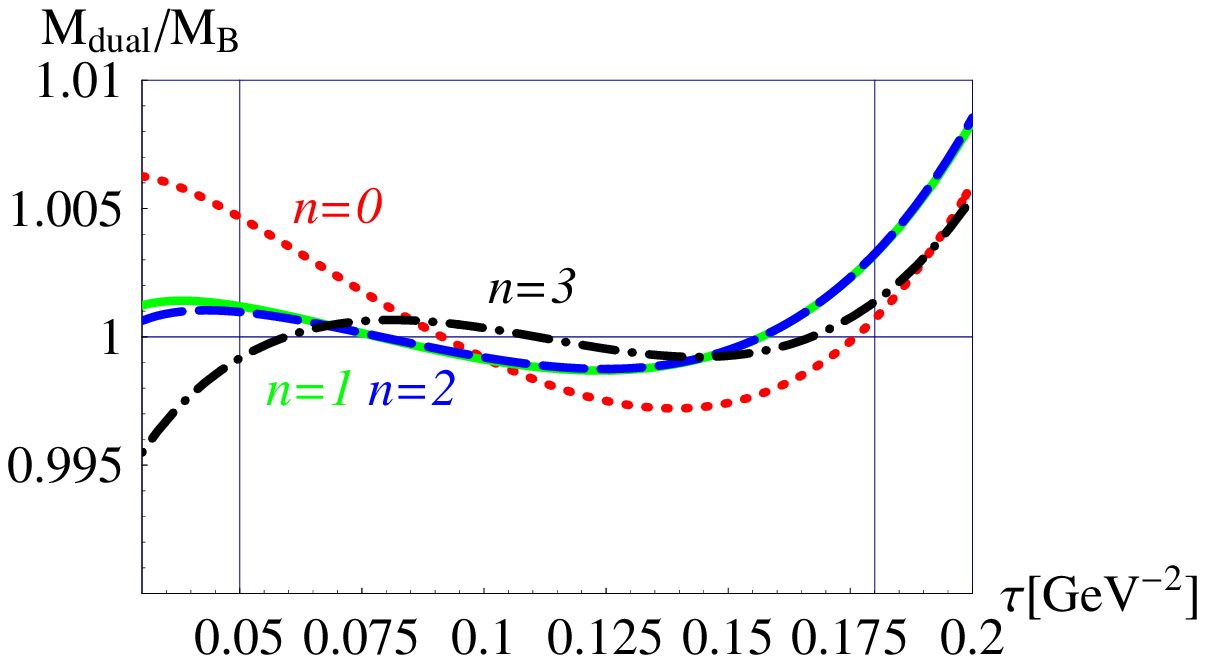}&
\includegraphics[width=6.7cm]{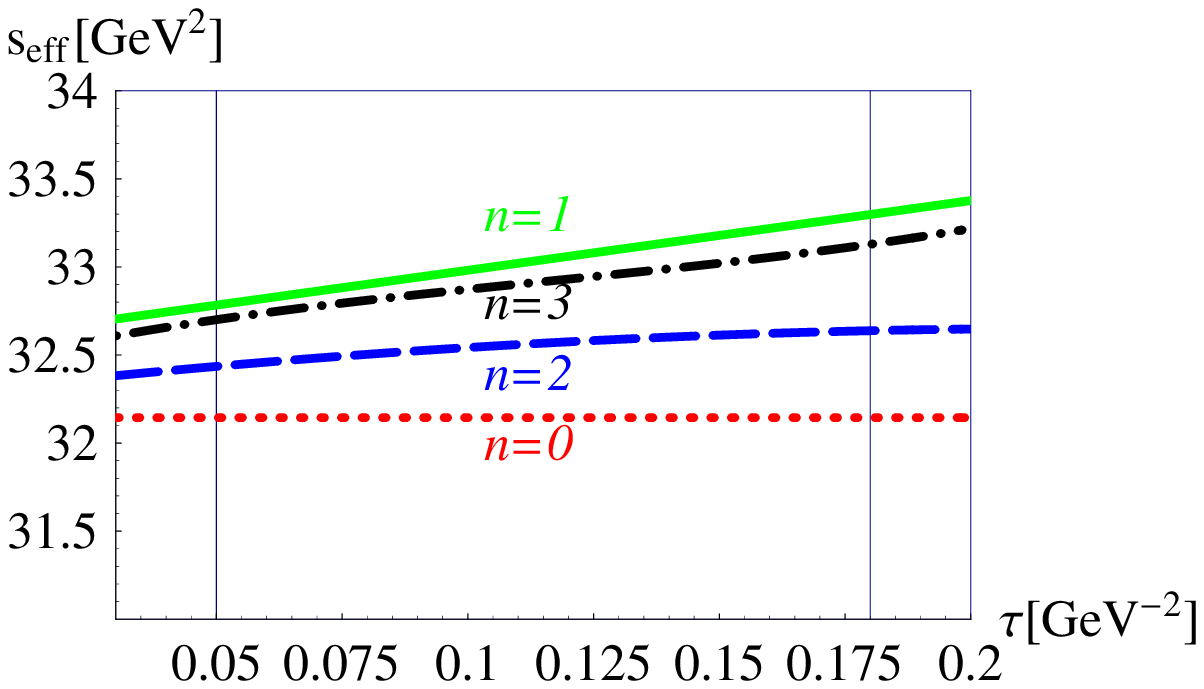}\\(a)&(b)\\
\includegraphics[width=6.7cm]{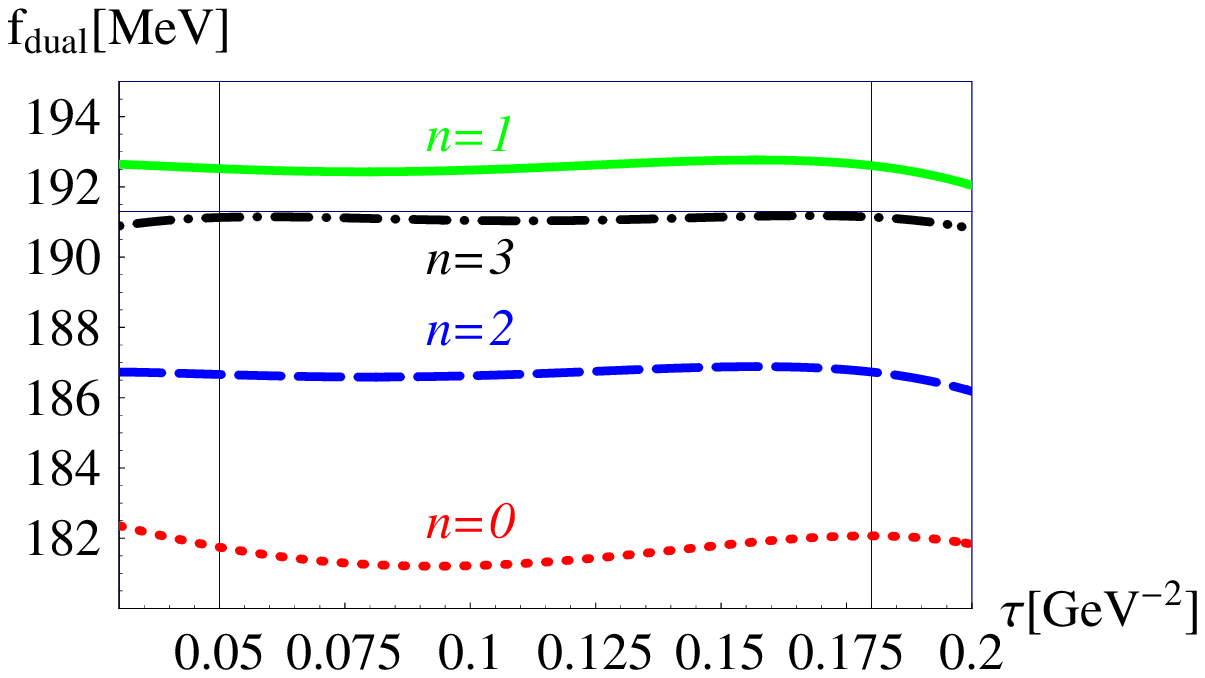}&
\includegraphics[width=6.7cm]{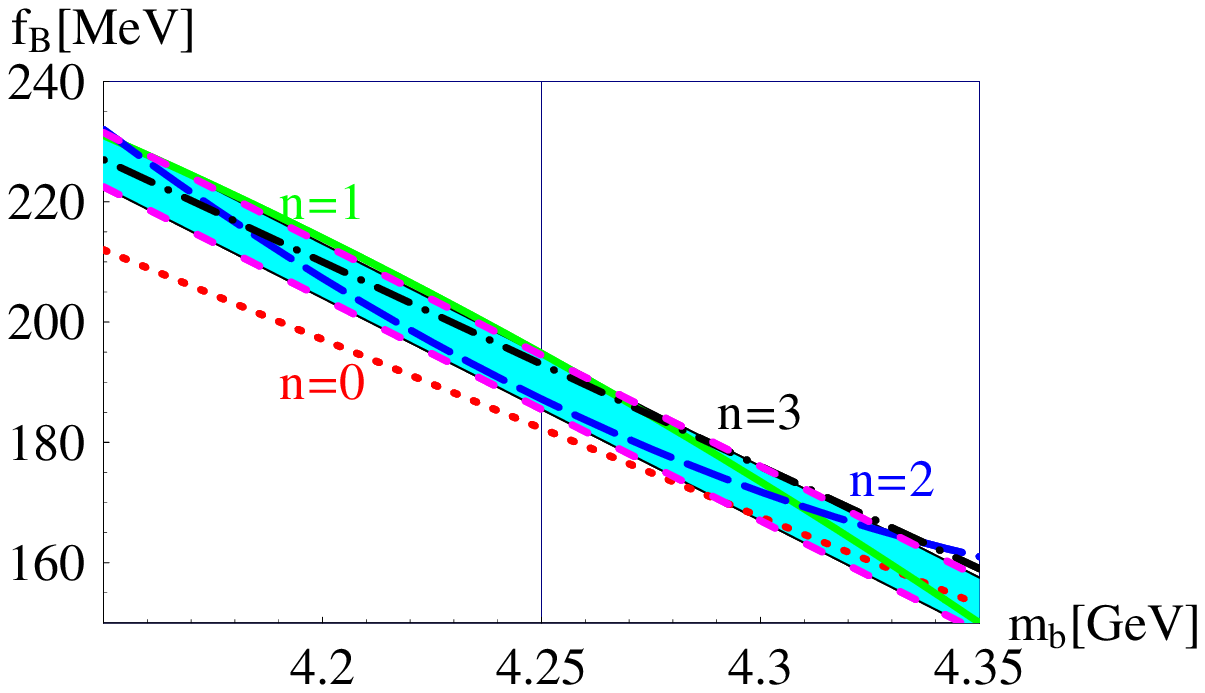}\\(c)&(d)
\end{tabular}
\caption{Dual mass $M_{\rm dual}(\tau)$ (a), corresponding
$\tau$-dependent effective continuum threshold $s_{\rm eff}(\tau)$
according to our Ansatz (\ref{zeff}), determined by minimizing the
expression (\ref{chisq}) (b), and dual decay constant $f_{\rm
dual}(\tau)$ (c). Results for
$m_b\equiv\overline{m}_b(\overline{m}_b)=4.25\;\mbox{GeV},$
$\mu=\nu=m_b,$ and central values of the other relevant parameters
are shown. (d) Dual decay constant of the $B$ meson vs.\ $m_b$~for
$\mu=\nu=m_b$ and central values of all the other OPE parameters.
The integer $n=0,1,2,3$ is the degree of the $s_{\rm eff}(\tau)$
polynomial in the Ansatz (\ref{zeff}). Red (dotted) line: $n=0$;
green (solid) line: $n=1$; blue (dashed) line: $n=2$; (black)
dot-dashed~line:~$n=3$.}\label{Plot:fB}\end{figure}

Recall that the
$\tau$-window for the $B_{(s)}$ mesons is fixed by the above
criteria to be equal to $\tau=(0.05$--$0.175)\;\mbox{GeV}^{-2}$.
Figure~\ref{Plot:fB}~shows~the corresponding results for the
effective continuum threshold $s_{\rm eff}(\tau)$ and the
extracted $f_B$. Obviously, in this window~the $\tau$-dependent
effective thresholds reproduce the meson mass $M_B$ much better
than the constant~one (Fig.~\ref{Plot:fB}a). This signals that
those dual correlators that correspond to such $\tau$-dependent
thresholds are less contaminated by the excited states.

According to Fig.~\ref{Plot:fB}d, the dependence of our QCD
sum-rule prediction for the $B$-meson decay constant $f_B$ on
$m_b$ and the quark condensate $\langle \bar qq\rangle\equiv
\langle \bar qq(2\;{\rm GeV})\rangle,$ for fixed values of the
other OPE parameters, may be well parameterized~by
\begin{equation}
f_B^{\rm dual}(m_b,\mu=\nu=m_b,\langle\bar qq\rangle)=\left[192.0
-37\left(\frac{m_b-4.247\;\mbox{GeV}}{0.1\;\mbox{GeV}}\right)
+4\left(\frac{|\langle\bar
qq\rangle|^{1/3}-0.269\;\mbox{GeV}}{0.01\;\mbox{GeV}}\right)
\pm3_{\rm(syst)}\right]\mbox{MeV},
\end{equation}
representing the range of results obtained for $n=1,2,3$ in the
Ansatz (\ref{chisq}) within the two short-dashed lines
in~Fig.~\ref{Plot:fB}d.

Note that our algorithm, relying on polynomial functions, provides
a clear and unambiguous prescription for fixing the effective
continuum thresholds. The $\tau$-dependence of the latter is
crucial for deriving the dual mass, the definition of which
involves a derivative w.r.t.~$\tau.$ On the other hand, our
decay-constant prediction may be reproduced by the constant
effective continuum threshold $s_{\rm
eff}=(33.1\pm0.5)\;\mbox{GeV}.$ However, in order to obtain this
very range of values, one has to apply our algorithm, which takes
advantage of the freedom provided by the $\tau$-dependence of the
thresholds.

Performing the bootstrap analysis of the OPE uncertainties and
adding the half-width of the band deduced from~our
$\tau$-dependent Ans\"atze~for the effective continuum threshold
of degree $n=1,2,3$ as (intrinsic) systematic error, we find
\begin{equation}
\label{fB}
f_B=\left(192.0\pm14.3_{\rm(OPE)}\pm3.0_{\rm(syst)}\right)\mbox{MeV}.
\end{equation}
The main contributions to the OPE uncertainty in the extracted
$f_B$ arise from the renormalization-scale dependence and the
errors in $m_b$ and the quark condensate. Let us emphasize that 
for $m_b=4.05\;\mbox{GeV}$ one gets
$f_B=265\;\mbox{MeV}$ which is very far from the result reported in
\cite{nar2001}, cf.~Table~\ref{Table:1}.

\subsection{\boldmath Decay constant of the $B_s$ meson}
A similar procedure yields for the $B_s$ meson
\begin{eqnarray}
f_{B_s}^{\rm dual}(m_b,\mu=\nu=m_b,\langle\bar ss\rangle)=\left[
228.0-43\left(\frac{m_b-4.247\;\mbox{GeV}}{0.1\;\mbox{GeV}}\right)
+3.5\left(\frac{|\langle\bar
ss\rangle|^{1/3}-0.248\;\mbox{GeV}}{0.01\;\mbox{GeV}}\right)
\pm4_{\rm(syst)}\right]\mbox{MeV}.
\end{eqnarray}
Performing the bootstrap analysis of the OPE uncertainties, we
obtain
\begin{eqnarray}
\label{fBs}
f_{B_s}=\left(228.0\pm19.4_{\rm(OPE)}\pm4_{\rm(syst)}\right)\mbox{MeV}.
\end{eqnarray}

\subsection{\boldmath $f_{B_s}/f_B$}
The resulting ratio of the $B$ and $B_s$ decay constants reads
\begin{eqnarray}
\label{ratioB}
f_{B_s}/f_B=1.184\pm0.023_{\rm(OPE)}\pm0.007_{\rm(syst)},
\end{eqnarray}
in excellent agreement with the recent lattice results summarized
in Table~\ref{table:lattice}. The error in the ratio
(\ref{ratioB}) arises mainly from the uncertainties in the quark
condensates $\langle\bar ss\rangle/\langle\bar
qq\rangle=0.8\pm0.3.$

\section{Extraction of the bottom-quark mass}The results of the
previous section reveal a strong sensitivity of the sum-rule
predictions for $f_B$ and $f_{B_s}$ on the~precise value of $m_b,$
in accordance with our simple quantum-mechanical analysis. This
feature opens the promising possibility to extract an accurate
value of the $b$-quark mass
$m_b\equiv\overline{m}_b(\overline{m}_b)$ by exploiting the
accurate lattice results for $f_B$ and~$f_{Bs}$.

\begin{figure}[htb]
\begin{tabular}{cc}
\includegraphics[width=8.cm]{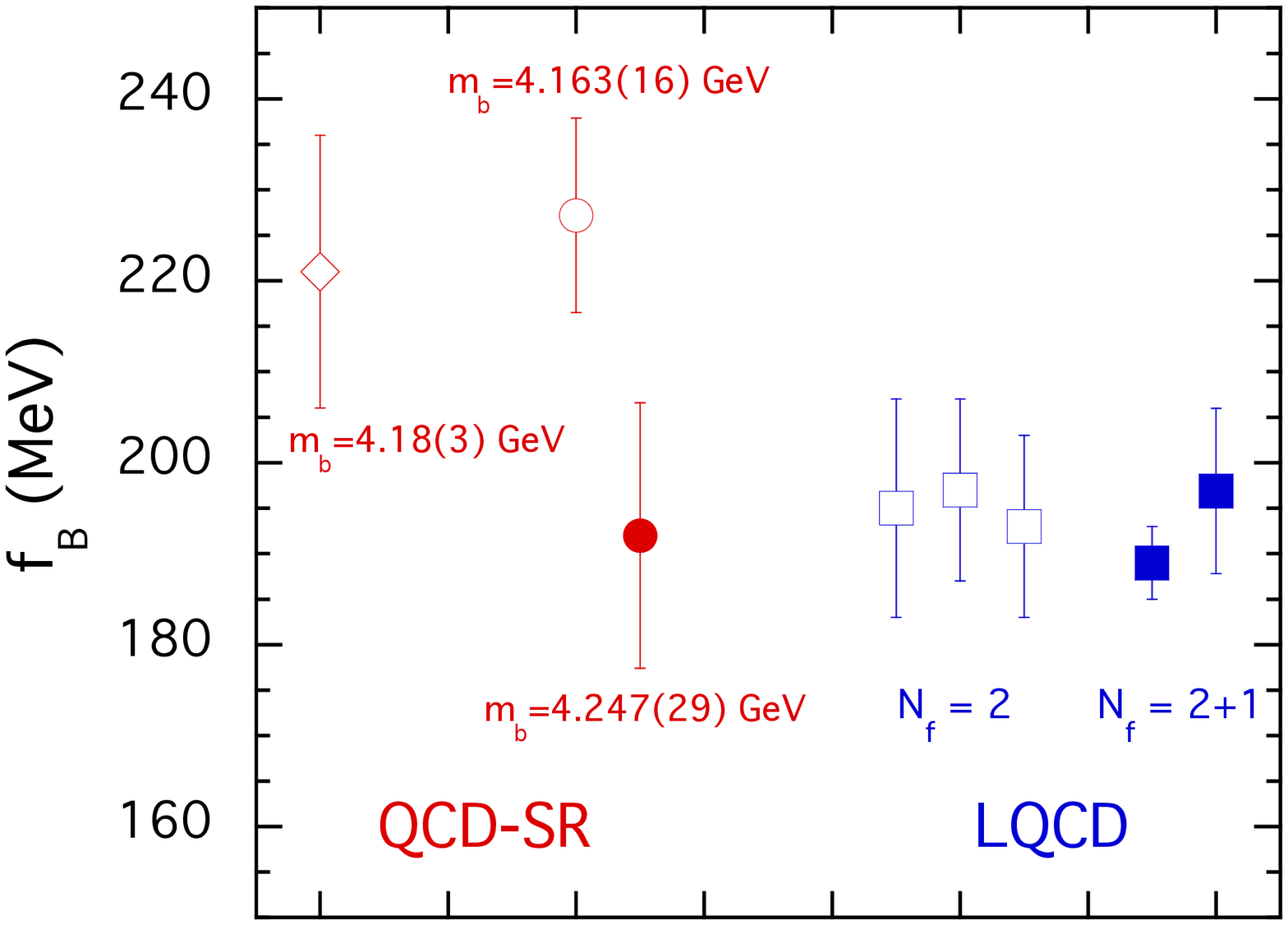}\hspace{5ex}&\hspace{5ex}
\includegraphics[width=8.cm]{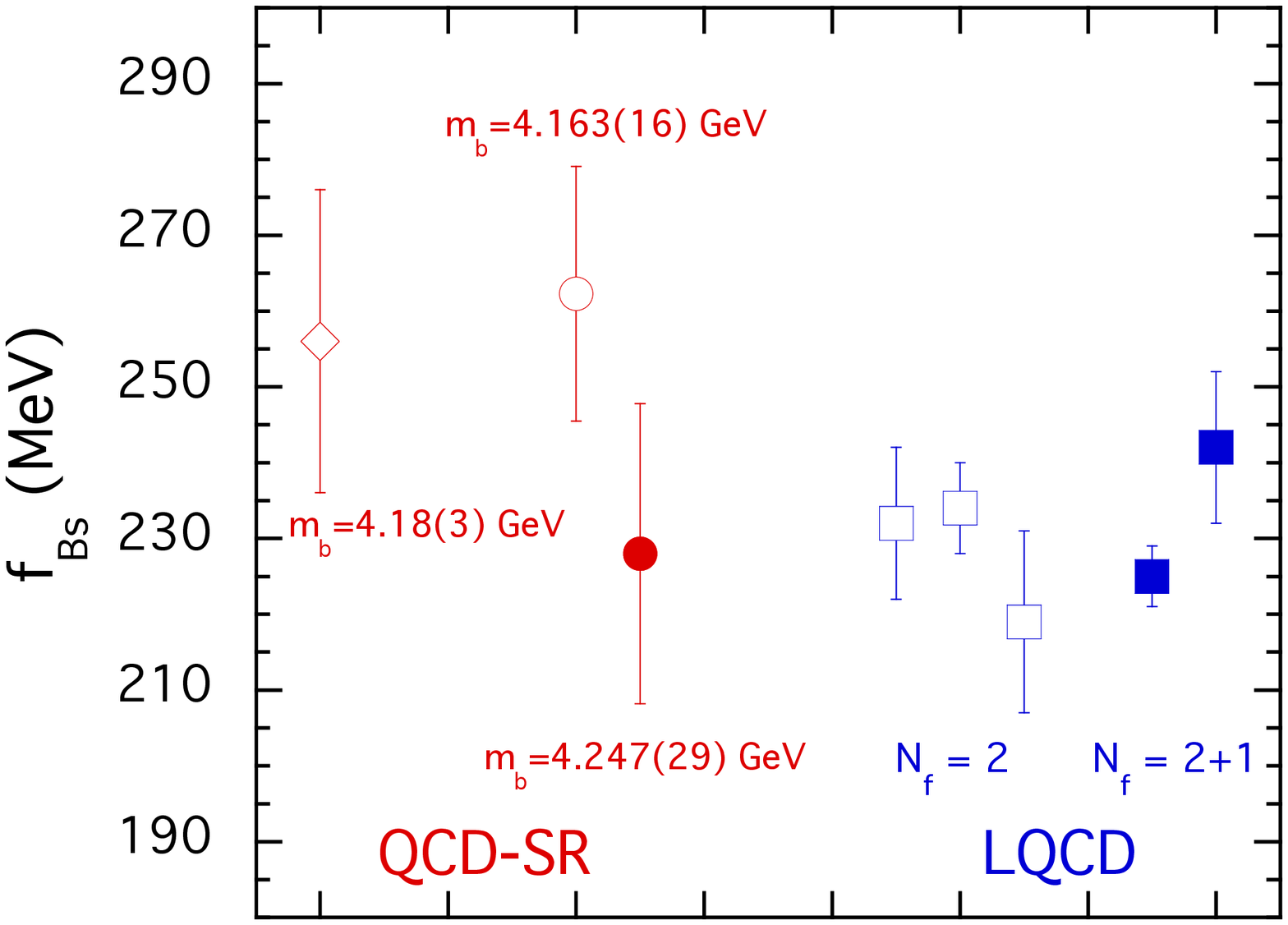}\\[1ex](a)&(b)\\[3ex]
\includegraphics[width=8.cm]{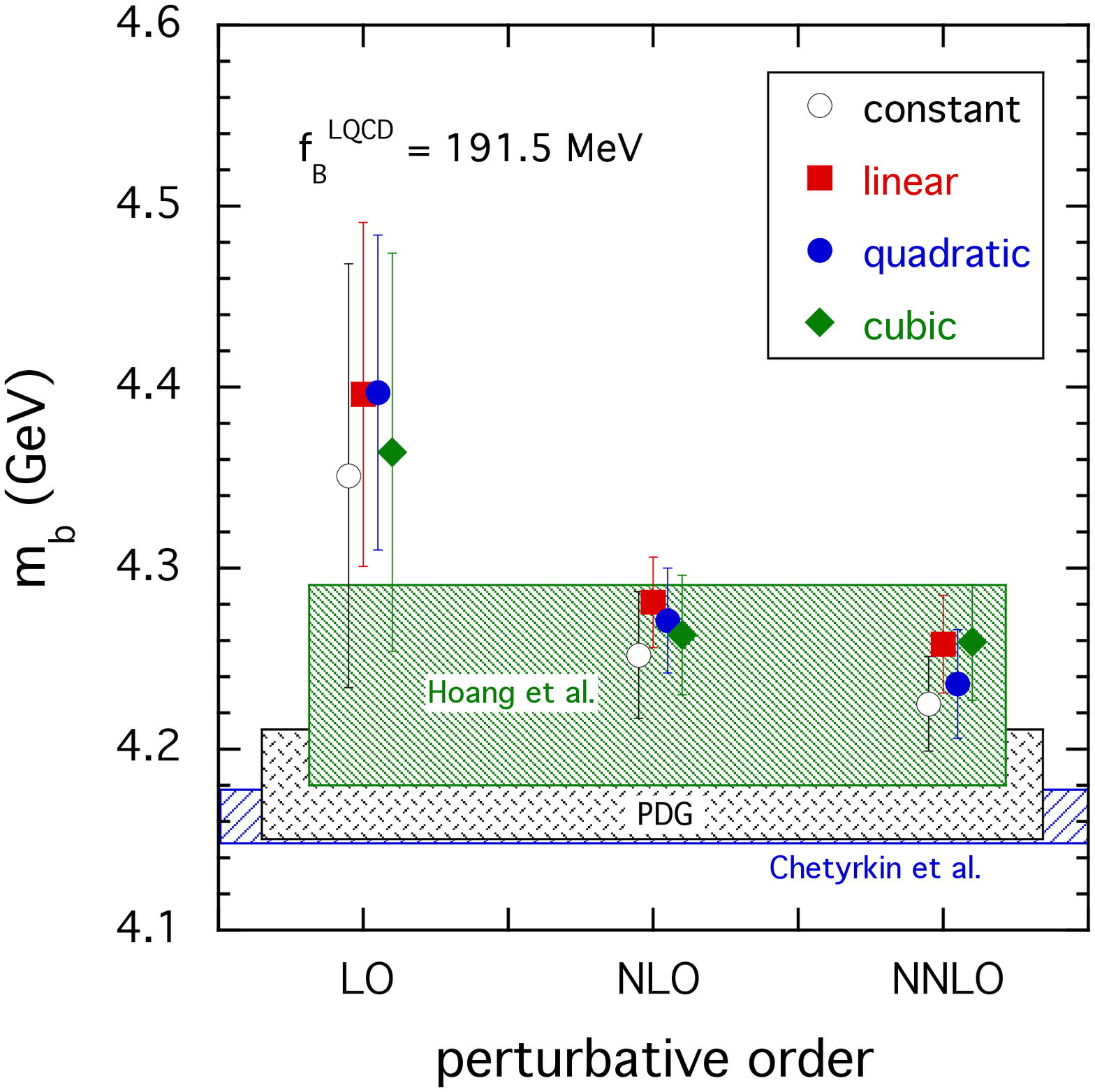}\hspace{5ex}&\hspace{5ex}
\includegraphics[width=8.cm]{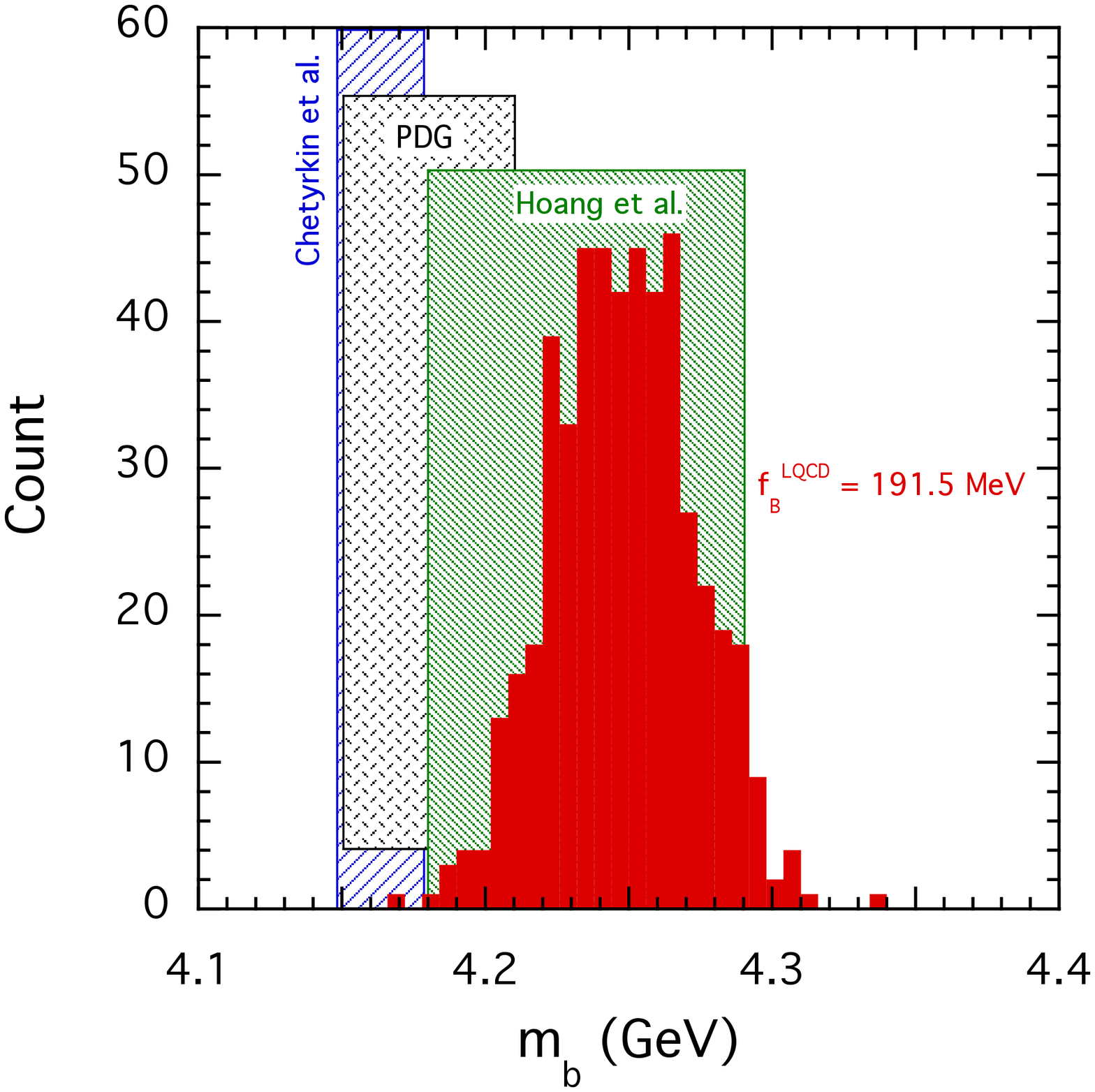}\\[1ex](c)&(d)
\end{tabular}
\caption{(a) Summary of our results for $f_B.$ Lattice (LQCD)
outcomes are from \cite{ETMC1,ETMC2,ALPHA} for two dynamical light
flavors ($N_f = 2$) and from \cite{HPQCD2,MILC} for
three dynamical flavors ($N_f=2+1$). For the $\tau$-dependent QCD
sum-rule (QCD-SR) result, the error shown~is the sum of the OPE
and systematic uncertainties in (\ref{fB}), added in quadrature.
(b) Similar findings for $f_{B_s}$. (c) The value of~$m_b$
extracted from the sum rule (\ref{sr}) by a bootstrap analysis of
the OPE uncertainties making use of the central value $f_B=191.5\;\mbox{MeV}$~and~the other relevant parameters
collected in (\ref{Table:2}). The dependence of $m_b$ on the
number of terms in the perturbative expansion of~the correlator is
indicated by LO, NLO, and NNLO. The shaded areas correspond to
$\pm1\sigma$ intervals of the results by PDG \cite{pdg}, 
Chetyrkin~et~al.\
\cite{mb1} and Hoang et al.\ \cite{hoang}. (d) Distribution of
$m_b$ as obtained by the bootstrap analysis described in the text. 
Gaussian distributions for all the OPE parameters (apart from the scales $\mu$ and $\nu$) 
with the associated uncertainties collected in (\ref{Table:2}) are employed. 
For the independent parameters $\mu$ and $\nu$, uniform distributions
in the range $3\;\mbox{GeV}<\mu,\nu< 6\;\mbox{GeV}$ are assumed.}
\label{Plot:mb}
\end{figure}

The latest lattice-QCD findings for these decay constants are recalled in Table \ref{table:lattice} 
and Fig.~\ref{Plot:mb}a,b (see also \cite{lattice}). Using these results and applying the
algorithms described above, the sum rule (\ref{sr}) yields the
results for $m_b$ shown in Fig.~\ref{Plot:mb}.
Figure \ref{Plot:mb}c presents the extracted values of $m_b$
depending on the number of terms kept in the perturbative part of
the correlator. Moving from $O(1)$ (LO) to $O(\alpha_s)$ (NLO)
accuracy of the perturbative expansion has two effects:~first,~the
central $m_b$ value decreases sizeably from 
$m_b^{\rm LO}=(4.38\pm0.1_{\rm(OPE)}\pm0.020_{\rm(syst)})\;\mbox{GeV}$ to 
$m_b^{\rm NLO}=(4.27\pm0.04_{\rm(OPE)}\pm0.015_{\rm(syst)})\;\mbox{GeV},$ 
and, second, the OPE-error also reduces considerably. Adding the
$O(\alpha_s^2)$ (NNLO) correction does not, however, entail a
sizeable~change of the predictions: 
$m_b^{\rm NNLO}=(4.247\pm0.027_{\rm(OPE)}\pm0.011_{\rm(syst)})\;\mbox{GeV}.$ 
Obviously, the extracted values of~$m_b$ exhibit a nice ``convergence''
depending on the accuracy of the perturbative correlation function.

The OPE error in the extracted $m_b^{\rm NNLO}$ is related to the 
variations of the OPE parameters in the ranges given in (\ref{Table:2}) and the independent 
variations of the scales $\mu$ and $\nu$ in the range 3 GeV $\le \mu,\nu \le$ 6 GeV. 
The individual contributions to the OPE error read: 
14 MeV ($\mu$, $\nu$), 20 MeV (quark condensate), 7 MeV (gluon condensate), 
8 MeV ($\alpha_s$), and 4 MeV (light-quark mass). 
Adding these values in the quadrature gives 27 MeV. 
The systematic uncertainty in the extracted value of $m_b$ is found 
as the spread of the~results for different Ans\"atze for the effective continuum threshold and amounts to 11 MeV.
To obtain the final estimate for $m_b$ one should further add the (Gaussian) error 18 MeV,  
related to the uncertainty in the lattice value of $f_B=(191.5\pm 7.3)$ MeV. 

The $O(\alpha_s^3)$ correction to the perturbative spectral density is, at present, not known. Nevertheless, on the
basis~of~our findings we do not expect a sizeable shift of the
central value of $m_b$ due to the inclusion of the $O(\alpha_s^3)$
correction. One, however, might expect a reduction of the
sensitivity of the extracted value of $m_b$ to the precise values
of the scales $\mu$ and $\nu$ and thus a further increase of the 
accuracy of the extracted value of the bottom-quark mass.

%*********************************************************************

\section{Summary and conclusions}
\label{Sec:conc}
We performed a detailed QCD sum-rule analysis of the $B$- and
$B_s$-meson decay constants, with particular emphasis on the study
of the errors in the extracted decay-constant values: the OPE
uncertainty due to the errors of the QCD parameters and the
intrinsic error of the sum-rule approach due to the limited
accuracy of the extraction procedure. Our main findings may be
summarized by the following observations:
\begin{enumerate}
\item
The choice of the renormalization scheme used to define the
heavy-quark mass is crucial for the convergence of the
perturbative expansion of the two-point function: the latter
exhibits in its pole-mass formulation no sign of convergence but
develops in its running-mass formulation a clear hierarchy of the
perturbative contributions. For the extracted decay constant, the
pole-mass result is sizeably smaller than its running-mass
counterpart,~albeit both enjoy a perfect stability in the Borel
parameter. Borel stability does not imply reliability of
sum-rule~results.
\item
The extraction of hadronic properties is
significantly improved by allowing a Borel-parameter dependence for the effective 
continuum threshold, which then quite naturally increases
the accuracy of the duality approximation. As shown already
before in the charmed-meson sector \cite{lms_fD}, considering
suitably optimized polynomial Ans\"atze for the effective
continuum threshold provides an estimate~of the intrinsic
uncertainty of the method of QCD sum rules.\item For beauty
mesons, a very strong correlation between the exact $m_b$ value
and the sum-rule result for $f_B$ is found:
\begin{eqnarray}
\frac{\delta f_B}{f_B}\approx-8\,\frac{\delta m_b}{m_b}.
\end{eqnarray}
This enables us to revert the problem and to make use of the
precise lattice-QCD computations of $f_B$ to extract the value of
$m_b$. Combining our sum-rule analysis with the latest results for
$f_B$ and $f_{Bs}$ from lattice QCD~yields
\begin{eqnarray}
\label{ourmb}
m_b=(4.247\pm0.027_{\rm (OPE)}\pm 0.018_{\rm (exp)}\pm0.011_{\rm(syst)})\;\mbox{GeV};
\end{eqnarray}
the OPE error is related to the uncertainties in the OPE input parameters, 
and the ``exp'' error is induced by the error in the lattice determination of $f_B$. 
Good news is that the systematic uncertainty of the sum-rule method, estimated
from the spread of the results for different Ans\"atze of the
effective continuum threshold, amounts to $11\;\mbox{MeV}$ and remains under control.
Adding all three error in quadrature yields our final estimate 
\begin{eqnarray}
\label{ourmb1}
m_b=(4.247\pm0.034)\;\mbox{GeV};
\end{eqnarray}
With (\ref{ourmb1}), the QCD sum rules for heavy--light correlators evaluated 
at $O(\alpha_s^2)$ accuracy yield, for the decay~constants,
\begin{align}
\label{fB_final}
f_B&=\left(192.0\pm14.3_{\rm(OPE)}\pm3.0_{\rm(syst)}\right)\mbox{MeV},\\
\label{fBs_final}
f_{B_s}&=\left(228.0\pm19.4_{\rm(OPE)}\pm4_{\rm(syst)}\right)\mbox{MeV},\\
\label{ratio_final}
f_{B_s}/f_B&=1.184\pm0.023_{\rm(OPE)}\pm0.007_{\rm (syst)}.
\end{align}
Our algorithm enables us to provide both the OPE uncertainties and
the intrinsic (systematic) uncertainty~of~the sum-rule method
related to the limited accuracy of the extraction procedure. We
observe an extreme sensitivity of the decay constant to the input
value of the quark mass, but only for beauty mesons. It is not
observed in~the charm sector, where one finds $\delta
f_D/f_D=-0.3\,\delta m_c/m_c$ \cite{lms_fD}. Therefore, the
extracted value of $f_D$ is rather~mildly sensitive to the precise
value of $m_c.$ In the charm sector, on the other hand, one
observes a stronger sensitivity~of the extracted value of $f_D$ to
the algorithm adopted for fixing the $\tau$-dependent effective
continuum threshold~\cite{lms_fD}.\item Our value (\ref{ourmb}) of
$m_b$ is extracted from the Borel QCD sum rule for the
heavy--light correlator known to $O(\alpha_s^2)$ accuracy. Taking
into account that the value of $m_b$ is changing only marginally
when moving from the $O(\alpha_s)$~to $O(\alpha_s^2)$ accuracy of
the correlator, we do not expect that the inclusion of the
presently unknown $O(\alpha_s^3)$ correction will lead to a
substantial change in the extracted value of $m_b$. Our result is
compatible with the result \cite{mb4}
\begin{eqnarray}
m_b=(4.209\pm0.050)\;\mbox{GeV}
\end{eqnarray}
found from moment sum rules for heavy--heavy correlators known to
the same $O(\alpha_s^2)$ accuracy as in our~analysis. We observe
an excellent agreement with the prediction of the RG-improved NNLL
analysis of the $\Upsilon$ sum rule~\cite{hoang},
\begin{eqnarray}
m_b=(4.235\pm0.055_{\rm(pert)}\pm0.003_{\rm(exp)})\;\mbox{GeV}.
\end{eqnarray}
Our result agrees within 2$\sigma$ with the PDG estimate  
\begin{eqnarray}
m_b=(4.18\pm0.03)\;\mbox{GeV}.
\end{eqnarray}
We realize, however, a pronounced tension with the predictions of \cite{mb1}
\begin{eqnarray}
m_b=(4.163\pm0.016)\;\mbox{GeV},
\end{eqnarray}
and \cite{dominguez}
\begin{eqnarray}
m_b=(4.171\pm0.009)\;\mbox{GeV},
\end{eqnarray}
based on sum rules for heavy--heavy correlators calculated
to $O(\alpha_s^3)$ accuracy. As already noticed above, it seems
unlikely that the $O(\alpha_s^3)$ correction may bring our result
in agreement with the relatively low value of~\cite{mb1};
therefore, we expect that this tension will persist. The origin of
this disagreement requires further~considerations.
\end{enumerate}
We conclude by emphasizing that the properly formulated Borel QCD
sum rules for heavy--light correlators provide a competitive tool
for the reliable calculation of heavy-meson properties and for the
extraction of basic QCD parameters by making use of the results
from lattice QCD and the experimental data. We point out that in
the context of QCD sum rules based on correlation functions
calculated at $O(\alpha_s^2)$ accuracy, Eq.~(\ref{ourmb}) gives the
appropriate value of the $b$-quark mass.

\vspace{.5cm}{\bf Acknowledgements.} 
We thank Andre Hoang for
helpful discussions. D.M.\ was supported by the Austrian Science
Fund (FWF), Project No.~P22843. The work was supported in part by
a grant for leading scientific schools~3920.2012.2, and by FASI
state Contract No.\ 02.740.11.0244.


\begin{thebibliography}{30} 
\bibitem{pdg}
J. Beringer {\em et al.} (Particle Data Group), Phys.~Rev.~D {\bf 86}, 010001 (2012).
\bibitem{ETMC1}
P.~Dimopoulos {\em et al.} (ETM Collaboration),
%``Lattice QCD determination of m_b, f_B and f_Bs with twisted mass Wilson fermions,''
JHEP {\bf 1201}, 046 (2012).%[arXiv:1107.1441 [hep-lat]].
\bibitem{ETMC2}
N.~Carrasco {\em et al.} (ETM Collaboration),
%, P.~Dimopoulos, R.~Frezzotti, V.~Gimenez, G.~Herdoiza, V.~Lubicz, G.~Martinelli and C.~Michael {\em et al.},
%``B-physics from the ratio method with Wilson twisted mass fermions,''
PoS(LATTICE 2012)104 (2012);
%``B-physics from lattice QCD...with a twist,''
PoS(ICHEP2012)428 (2012).
%[arXiv:1212.0301 [hep-ph]].
\bibitem{mb_Gimenez}
V.~Gimenez, L.~Giusti, G.~Martinelli, and F.~Rapuano,
%"NNLO unquenched calculation of the b quark mass,"
JHEP {\bf 03}, 018 (2000).
\bibitem{mb_UKQCD}
C.~McNeile, C.~Michael, and G.~Thompson (UKQCD Collaboration),
%"An unquenched lattice QCD calculation of the mass of the bottom quark,"
Phys.~Lett.~B {\bf 600}, 77 (2004).
\bibitem{ALPHA}
F.~Bernardoni {\em et al.} (ALPHA Collaboration),
Nucl.~Phys.~Proc.~Suppl.~{\bf 234}, 181 (2013).
\bibitem{mb4}% 3-loops
J.~H.~K\"uhn and M.~Steinhauser, Nucl.~Phys.~B {\bf 619}, 588
(2001).
\bibitem{mb1}% 4-loops
K.~G.~Chetyrkin {\em et al.}, Phys.~Rev.~D {\bf 80}, 074010
(2009).
\bibitem{mb2}% 4-loops
C.~McNeile {\em et al.} (HPQCD Collaboration), Phys.~Rev.~D {\bf
82}, 034512 (2010).
\bibitem{hoang}
A.~Hoang, P.~Ruiz-Femenia, and M.~Stahlhofen, JHEP {\bf 10}, 188
(2012).
\bibitem{lms_qcdvsqm}W.~Lucha, D.~Melikhov, and S.~Simula,
Phys.~Lett.~B {\bf 687}, 48 (2010); Phys.~Atom.~Nucl.~{\bf 73},
1770 (2010).
\bibitem{neubert}M.~Neubert, Phys.~Rept.~{\bf 245}, 259 (1994).
\bibitem{svz}M.~A.~Shifman, A.~I.~Vainshtein, and V.~I.~Zakharov,
Nucl.~Phys.~B {\bf 147}, 385 (1979).
\bibitem{aliev}T.~M.~Aliev and V.~L.~Eletsky, Yad.~Fiz.~{\bf 38},
1537 (1983).
\bibitem{nar2001}S.~Narison, Phys.~Lett.~B {\bf 520}, 115 (2001).
\bibitem{jamin}M.~Jamin and B.~O.~Lange, Phys.~Rev.~D {\bf 65},
056005 (2002).
\bibitem{lms_fB}W.~Lucha, D.~Melikhov, and S.~Simula, J.~Phys.~G {\bf 38}, 105002 (2011).
\bibitem{nar2012}S.~Narison, Phys.~Lett.~B {\bf 718}, 1321 (2013).
\bibitem{chetyrkin}K.~G.~Chetyrkin and M.~Steinhauser,
Phys.~Lett.~B {\bf 502}, 104 (2001); Eur.~Phys.~J.~C {\bf 21}, 319 (2001).

\bibitem{HPQCD1}H.~Na {\em et al.},
%, C.~J.~Monahan, C.~T.~H.~Davies, R.~Horgan, G.~P.~Lepage and J.~Shigemitsu,
%``The B and B_s Meson Decay Constants from Lattice QCD,''
Phys.~Rev.~D {\bf 86}, 034506 (2012).
%[arXiv:1202.4914 [hep-lat]].
\bibitem{HPQCD2}C.~McNeile {\em et al.},
%, C.~T.~H.~Davies, E.~Follana, K.~Hornbostel and G.~P.~Lepage,
%``High-Precision $f_{B_s}$ and HQET from Relativistic Lattice QCD,''
Phys.~Rev.~D {\bf 85}, 031503 (2012).
%[arXiv:1110.4510 [hep-lat]].
\bibitem{MILC}A.~Bazavov {\em et al.} (Fermilab Lattice and MILC Collaborations),
%``B- and D-meson decay constants from three-flavor lattice QCD,''
Phys.~Rev.~D {\bf 85}, 114506 (2012).
% [arXiv:1112.3051 [hep-lat]].

\bibitem{lms_1}
W.~Lucha, D.~Melikhov, and S.~Simula, Phys.~Rev.~D
{\bf 76}, 036002 (2007); Phys.~Lett.~B {\bf 657}, 148 (2007);
Phys.~Atom.~Nucl.\ {\bf 71}, 1461 (2008); Phys.~Lett.~B {\bf 671},
445 (2009); D.~Melikhov, Phys.~Lett.~B {\bf 671}, 450 (2009).
\bibitem{lms_new}
W.~Lucha, D.~Melikhov, and S.~Simula, Phys.~Rev.~D
{\bf 79}, 096011 (2009); J.~Phys.~G {\bf 37}, 035003 (2010);
W.~Lucha, D.\ Melikhov, H.~Sazdjian, and S.~Simula, Phys.~Rev.~D
{\bf 80}, 114028 (2009).
\bibitem{Mb3loop}
K.~G.~Chetyrkin and M.~Steinhauser,
Phys.~Rev.~Lett.~{\bf 83}, 4001 (1999); Nucl.~Phys.~B {\bf 573},
617 (2000); K.~Melnikov and~T.\ van Ritbergen, Phys.~Lett.~B {\bf 482}, 99 (2000).
\bibitem{colangelo}
G.~Colangelo {\em et al.},
%``Review of lattice results concerning low energy particle physics,''
Eur.~Phys.~J.~C~{\bf 71}, 1695 (2011).
%[arXiv:1011.4408 [hep-lat]].
\bibitem{bootstrap}B.~Efron and R.~J.~Tibshirani, {\em An
Introduction to the Bootstrap}, Monographs on Statistics and
Applied Probability,~Vol.~{\bf 57}, CRC Press, 1993.
\bibitem{lms_fD}W.~Lucha, D.~Melikhov, and S.~Simula, Phys.~Lett.~B {\bf 701}, 82 (2011).

\bibitem{lattice}E.~Gamiz,
%''Flavour physics from lattice QCD'', Xth Quark Confinement and the hadron spectrum
PoS(Confinement X)241 (2012).%arXiv:1303.3971 [hep-ph]
\bibitem{dominguez}
S.~Bodenstein {\em et al.}, Phys.~Rev.~D {\bf 85}, 034003 (2012).
\end{thebibliography}
\end{document}